# To Heat or not to Heat: a Study of the Performances of Iron Carbide Nanoparticles in Magnetic Heating

*Juan Manuel Asensio,[a]\* Julien Marbaix,[a] Nicolas Mille,[a] Lise-Marie Lacroix,[a] Katerina Soulantica,[a] Pier-Francesco Fazzini,[a] Julian Carrey,[a] Bruno Chaudret.[a]\**

[a]LPCNO, Université de Toulouse, CNRS, INSA, UPS, 135 avenue de Rangueil, 31077 Toulouse, France.
\*Dr. Bruno Chaudret: chaudret@insa-toulouse.fr
\*Dr Juan M. Asensio: asensior@ insa-toulouse.fr

ABSTRACT. Heating magnetic nanoparticles with high frequency magnetic fields is a topic of interest for biological applications (magnetic hyperthermia) as well as for heterogeneous catalysis. This study shows why FeC NPs of similar structures and static magnetic properties display radically different heating power (SAR from 0 to 2 kW.g$^{-1}$). By combining results from Transmission Electron Microscopy (TEM), Dynamic Light Scattering (DLS) and static and time-dependent high-frequency magnetic measurements, we propose a model describing the heating mechanism in FeC nanoparticles. Using, for the first time, time-dependent high-frequency hysteresis loop measurements, it is shown that in the samples displaying the larger heating powers, the hysteresis is strongly time dependent. More precisely, the hysteresis area increases by a factor 10 on a timescale of a few tens of seconds. This effect is directly related to the ability of the nanoparticles to form chains under magnetic excitation, which depends on the presence or not of strong dipolar couplings. These differences are due to different ligand concentrations on the surface of the particles. As a result, this study allows the design of a scalable synthesis of nanomaterials displaying a controllable and reproducible SAR.

INTRODUCTION

Magnetic heating, i.e. heating of magnetic nanoparticles using magnetic excitation, is a topic of interest that proposes promising applications in the field of cancer therapy[1-3] –the so-called magnetic hyperthermia– and more recently in heterogeneous catalysis.[4-11] Thus, recent papers and patents evidence the high potential of this technique in catalysis given its anticipated energetic efficiency. A recurrent problem in magnetic hyperthermia concerns the variability of the heating capacities of nanomaterials and the origin of this phenomenon. The use of magnetic heating in catalysis requires materials displaying the highest possible heating powers in a reproducible way and therefore, a good understanding of the heating mechanism of magnetic nanoparticles (MNPs) and their assemblies under operating conditions. Magnetic heating has developed considerably during the past few years thanks nanochemistry, which has enabled the controlled synthesis of MNPs displaying tailored heating properties. However, controlling the heating power of MNPs in reproducible and scaled-up processes is a challenging task as the magnetic properties depend on many parameters.[3, 12] As a consequence, variations in the reaction conditions may lead to variations in the heating power of the MNPs, which emphasizes the need for robust synthetic methods for MNPs production.

The heating power of MNPs is usually quantified by the specific absorption rate (SAR), which describes the amount of energy absorbed per unit of mass in the presence of an alternating magnetic field. Although the mechanism of heating for single-domain NPs is well-understood,[13, 14] several theoretical and experimental studies have demonstrated that magnetic interactions between the NPs can dramatically affect their heating power.[15-21] A good understanding of the influence of interactions between the NPs on their SAR is, therefore, a prerequisite to any development of the use of magnetic heating whether in biology or in catalysis. Some theoretical and experimental works have shown that organization of NPs into chains as a result of the application of a magnetic field during the experiments, induces the occurrence of an additional uniaxial anisotropy that enhances their heating power.[19, 22-24] The magnitude of magnetic interactions (dipolar couplings) between the NPs is crucial to allow this phenomenon taking place, since if the NPs are too far away to interact and to be organized in the presence of the magnetic field, their heating power remains low. Thus, the heating power increases with the concentration of NPs until reaching an optimal point, after which higher concentrations favor the presence of strong demagnetizing interactions resulting into a drop of the heating power.[20]

One additional consequence of dipolar interactions is that, at a given concentration, an agglomeration of MNPs reduces their heating due to an absence of mobility[25] and/or to the fact that the overall magnetic moment of agglomerates is reduced due to their magnetic flux closure configurations, reducing their tendency to interact with their neighbors.[26, 27] Thus, in samples without agglomerates and displaying large heating powers, the heating power is expected to increase with time following the dynamics of the NPs self-organization upon application of an alternating magnetic field. On the contrary, in samples with agglomerates, heating power is expected to remain small and constant with time. However, to the best of our knowledge, direct experimental proofs of these expected effects have never been reported. This will be done in the present work and monitored by a technique never used in this context: time-dependent high-frequency measurements.

Different MNPs have been synthesized for applications in magnetic hyperthermia. Among them, the most widely used are iron oxide-based systems due to their good biocompatibility.[26, 28-33] However, for magnetically-induced catalysis where higher temperatures are required, iron carbide NPs (FeC NPs) can be better candidates. especially since these materials are ferromagnetic at room temperature and display moderate coercive fields and good colloidal stability.[2] Different preparation methods of FeCx NPs for applications in catalysis have been reported.[34-39] Over the past years, our group has developed the synthesis of monodisperse Fe(0) NPs, either pure or bimetallic[40-43] and their transformation into monodisperse iron carbide

NPs.[5, 34, 44] High values of SAR of ca. 750 W·g$^{-1}$ (100 kHz, 47 mT) were achieved for Fe(0) NPs,[45] but their transformation into FeC NPs led to much enhanced SAR as well as better catalytic activity and air stability. Indeed, we recently reported the synthesis of FeC$_x$ NPs displaying heating properties in organic solvents that surpass those of any previously described material.[5, 44] The NPs were synthesized by carbidization of pre-formed monodisperse Fe(0) NPs using a mixture of *syngas* (CO/H$_2$). We demonstrated that their very high SAR was related to the presence of the crystalline Fe$_{2.2}$C phase.[5] However, different values of SAR could be obtained upon slightly modifying the reaction and purification conditions.

In this paper, we demonstrate that standard static magnetic measurements cannot discriminate between heating and non-heating Fe$_{2.2}$C NPs. In our case, the heating properties can only be understood in terms of the presence or not of strong dipolar couplings resulting from the organization or agglomeration of the nanoparticles, which in turn depend on the surface ligands concentration. Time-dependent high-frequency magnetic measurements were employed for the first time to characterize the behavior of NPs assemblies under alternating magnetic fields. In association with dynamic light scattering they were found to be the key techniques to unveil the origin of the differences in the heating properties of the samples. These findings constitute a novel insight into the relationship between the synthesis of NPs and their final heating properties and allowed to design robust procedures for the production of FeC NPs based nanomaterials displaying high and reproducible heating powers.

RESULTS AND DISCUSSION

*Heating and non-heating Fe$_{2.2}$C NPs obtained by carbidization of Fe(0) NPs.*

In our previous work, we explored the synthesis of Fe$_{2.2}$C NPs through carbidization of pre-formed monodisperse Fe(0) NPs under a *syngas* mixture in mesitylene.[5] In that preliminary study, the carbidization was performed starting from Fe(0) NPs of 12.5 nm. The Fe(0) NPs had been isolated by post-synthesis decantation and washing, after which a powder with a Fe content of ca. 50 wt% was obtained and carbidized during 96 h to give $Fe_{2.2}C$ (**FeC-ref**) NPs with unprecedent SAR (see Table 1). The excellent heating power of the $Fe_{2.2}C$ NPs obtained after carbidization incited us to optimize and scale-up the reaction. Thus, in order to purify more extensively the surface of the particles, a more efficient washing of the powder led to Fe(0) NPs **Fe-1**, with a Fe content of 78% determined by thermogravimetric analysis (TGA). Surprisingly, even though the carbidization of these NPs was performed under similar conditions to those described in our previous work (reaction time of 96 hours), a strong agglomeration of the FeC NPs (NPs **FeC-1** in Table 1, see also Figure 1-a and S1) was observed by transmission electron microscopy (TEM).

The SAR of **FeC-1** measured by calorimetry experiments had a value of 0 W·g$^{-1}$ when applying an alternating magnetic field of $\mu_0 H_{rms}$ of 47 mT at a fixed frequency of 93 kHz, ($H_{max} = \sqrt{2} H_{rms}$) contrasting with the excellent SAR found under the same conditions for NPs **FeC-ref** prepared by the same methodology (ca. 2000 W·g$^{-1}$, see Table 1). In addition, NPs **FeC-1** were not dispersible in mesitylene, in contrast to the NPs **FeC-ref** prepared in our previous work. Furthermore, the NPs rapidly precipitated even after external addition of solubilizing ligands and sonication at room temperature. Finally, the size of the FeC NPs **FeC-1** after carbidization was only marginally increased compared to the starting Fe(0) NPs (from 12.5 ± 1.3 nm to 13.7 ± 1.6 nm). Extending the carbidization time to 7 or 8 days induced the coalescence of the NPs to give large nanoobjects of more than 200 nm observed by TEM. These results suggest a non-innocent role of the ligands in the carbidization process.

Analysis of the Mössbauer spectrum (see Fig S2) showed the presence of the heating carbide phase $Fe_{2.2}C$ in **FeC-1** with a concentration of 59.6%, together with some $Fe_5C_2$ (28.1%) and Fe(0) (8.3%). Although the relative amount of the heating phase $Fe_{2.2}C$ was lower than in **FeC-ref** NPs (71.2% of $Fe_{2.2}C$ in the latter), the absence of any heating capacities in **FeC-1** cannot be justified by this difference. Powder X-Ray diffraction (XRD) analysis of **FeC-1** only showed the peaks related to the Fe$_{2.2}$C phase (see Fig. S3). Similarly, the saturation magnetization (*M*s, 170 A·m$^2$·kg$^{-1}$), remnant magnetization (*Mr*, 23 A·m$^2$·kg$^{-1}$), coercive field (Hc, 0.051 T) and magnetic susceptibility (χ*m*, 0.043) of the **FeC-1** NPs measured by vibrating sample magnetometry (VSM, see Fig S4) did not reveal any significant differences in the *Ms*, *Mr* or χ*m* from the values found in the **FeC-ref** NPs previously reported by Bordet et al., although a higher value of Hc was obtained in that case (*M*s = 170 A·m2·kg-1, *Mr* ~ 30 A·m2·kg-1, Hc = 0.103 T and χ*m* ~ 0.04 at 300 K). All these results indicate that these structural differences are not enough to explain the lack of heating properties in **FeC-1** NPs.

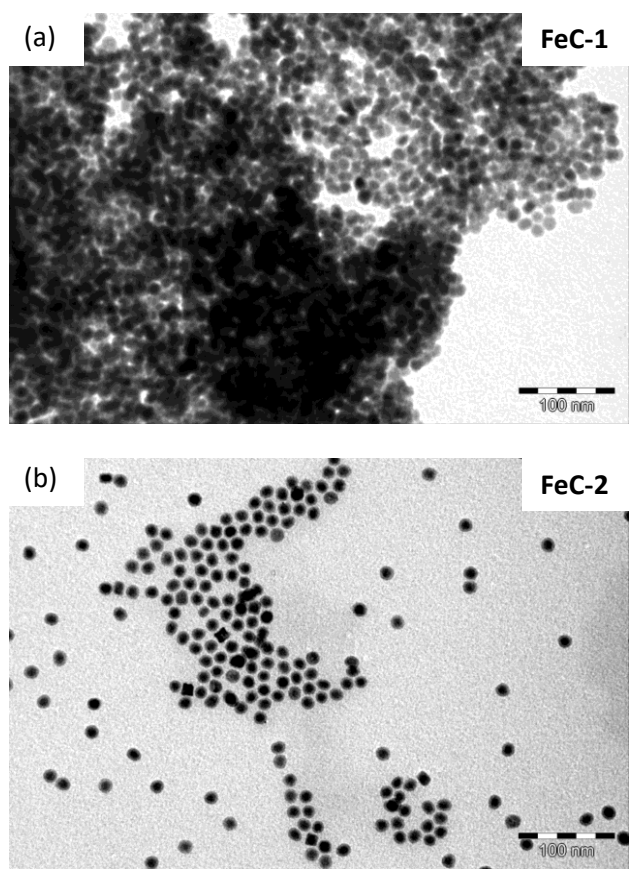

**Figure 1.** TEM micrographs of (a) the agglomerated NPs **FeC-1** after carbidization of 96 h and (b) of the dispersible NPs **FeC-2** after carbidization of 48 h.

High Resolution TEM (HRTEM) and High-Angle Annular Dark-Field Scanning Transmission Electron microscopy (HAADF-STEM) analyses of the non-heating **FeC-1** NPs showed that the agglomerates formed by several NPs are surrounded by a thin amorphous layer composed of a light element (see Figure S5). Energy-dispersive X-ray EDX spectroscopy also evidenced the presence of a shell containing silicon in addition to oxygen (see Figure 2b-f). This amorphous silicon-oxygen embedding layer is likely to be amorphous silica or a polysiloxane derivative, the presence of which could be related to the lack of solubility of NPs **FeC-1** in mesitylene and their assembly into the large aggregates as it would reduce the interactions between the agglomerates and the ligands added to re-disperse the NPs. XPS spectroscopy confirmed the presence of Si at binding energies of 102 eV for the Si $2p_{3/2}$ orbital, which are characteristic of silicates or organosiloxanes (see Fig S6). Concerning its origin, the only possible source of Si in the carbidization process is hexamethyldisilazane (*hmds*) resulting from the reaction with hydrogen of the corresponding amido ligand present in $\{Fe[N(SiMe_3)_2]_2\}_2$, the organometallic precursor of the starting Fe NPs (see experimental section for more details). When carbidization progresses, water molecules are released into the solution as a reaction product[44] and at low concentration of other ligands they can react with the Si moiety, leading to the formation of the amorphous embedding layer.

The role of this layer in the *on-off* behaviour of the heating power has then to be clarified. It has been previously shown that the presence of more than 2 wt% of $SiO_2$ deteriorates the magnetic permeability of a *hard* magnetic material such as bulk barium ferrite,[46] but here, this is not the case since the *Ms* of the **FeC-1** NPs determined by VSM is very similar to that measured for the **FeC-ref** NPs (see Table 1). Then, the *on-off* behaviour of the heating power has to be related to the NPs agglomeration, in which the silicon containing layer can also play a role by preventing chain formation. There are two convergent explanations to this fact: (i) when the NPs are aggregated, the proximity between the magnetic domains induces strong demagnetizing interactions between them, typically dipole-dipole interactions[18, 20, 47] and (ii), when the FeC NPs are agglomerated, their organization in the magnetic field into chains is not possible. Such chains are necessary to be formed during the hyperthermia experiments in order to induce the presence of a uniaxial anisotropy that enhances the NP heating power.[19, 22] For this phenomenon to happen, a relative free mobility of the NPs is required.

We noticed that after 96 hours of carbidization, 2 bars of *syngas* were consumed for the preparation of the **FeC-1** NPs, whereas for the synthesis of the heating **FeC-ref** NPs the pressure drop observed at the end of the reaction was of only 0.8 bars. We assume that due to the presence of less organic ligands in the **Fe-1** NPs herein used, an easier surface access may accelerate CO hydrogenation, leading to more *syngas* consumption and justifying the higher pressure drop. When carbidization of **Fe-1** NPs was allowed to proceed only for 48 hours, the *syngas* consumption was of 0.8 bars, and the **FeC-2** NPs obtained were dispersible in mesitylene as evidenced by TEM micrographs (see Fig 1b and Fig S7). In contrast to **FeC-1**, their SAR measured at $\mu_0H_{rms}$ of 47 mT with a $f$ of 93 kHz was 1260 W.g$^{-1}$ (see Table 1 and Fig S8). The **FeC-2** NPs were also characterized by the same techniques as the **FeC-1** NPs (see Fig S9-11). The SAR of NPs **FeC-2** was higher than the SAR of the analogous NPs obtained by Bordet et al. after 48 hours of carbidization (1260 and 342 W·g$^{-1}$ respectively).

**Table 1.** Comparison between the NPs **FeC-1** and **FeC-2** prepared in this work by variation of the reaction time with the NPs **FeC-ref** prepared by Bordet et al.[5]

| Carbidization conditions: | | | | |
|---|---|---|---|---|
| | NPs **FeC-1** (this work, 96 h carbidization) | NPs **FeC-2** (this work, 48 h carbidization) | NPs **FeC-ref** (Bordet et al., 96 h carbidization) | NPs **FeC-ref** (Bordet et al., 48 h carbidization) |
| Size (nm) | 13.3 ± 1.6 nm | 14.4 ± 1.3 nm | 15.9 ± 0.9 nm | ~ 13 nm |
| ΔP after reaction (bar) | >2 | 0.8 | ~0.8 | ~0.4 |
| Starting Fe(0) NPs (Fe wt%[a]) | **Fe-1** (78%) | **Fe-1** (78%) | **Fe-ref** (ca. 50%) | **Fe-ref** (ca. 50%) |
| Dispersible in Mesitylene | No | Yes | Yes | Yes |
| SAR (W·g$^{-1}$)[b] | 0 | 1260 | 1935 | 342 |
| Magnetic properties[c] | $M_s$ = 170 A·m$^2$·kg$^{-1}$<br>$M_r$ = 23 A·m$^2$·kg$^{-1}$<br>$H_c$ = 0.051 T<br>$\chi_m$ = 0.043 | $M_s$ = 168 A·m$^2$·kg$^{-1}$<br>$M_r$ = 28 A·m$^2$·kg$^{-1}$<br>$H_c$ = 0.055 T<br>$\chi_m$ = 0.043 | $M_s$ = 170 A·m$^2$·kg$^{-1}$<br>$M_r$ ~ 30 A·m$^2$·kg$^{-1}$<br>$H_c$ = 0.103 T<br>$\chi_m$ ~ 0.04 | $M_s$ = 163 A·m$^2$·kg$^{-1}$<br>$M_r$ ~ 25 A·m$^2$·kg$^{-1}$<br>$H_c$ = 0.050 T<br>$\chi_m$ ~ 0.04 |
| Mössbauer spectroscopy[d] | Fe$_{2.2}$C (60 %)<br>Fe$_5$C$_2$ (28%)<br>Fe(0) (8%) | Fe$_{2.2}$C (52%)<br>Fe$_5$C$_2$ (32%)<br>Fe(0) (12%) | Fe$_{2.2}$C (71%)<br>Fe$_5$C$_2$ (24%)<br>Fe(0) (4%) | Fe$_{2.2}$C (55%)<br>Fe$_5$C$_2$ (23%)<br>Fe(0) (18%) |

[a] Deterimined by TGA. [b] Measured at 47 mT and 93 kHz. [c] Determined by VSM and measured at 300 K. [d] Measured at 4 K.

Interestingly, despite their better heating properties, the degree of carbidization of the **FeC-2** NPs was lower than for **FeC-1**, as deduced from the lower content of the Fe$_{2.2}$C and Fe$_5$C$_2$ phases in the Mössbauer spectroscopy (see Table 1). This fact also indicates that the absence of heating power in NPs **FeC-1** is not related to the carbidization degree and raises the question of the heating mechanism. From the above data we can conclude that the low ligand content of the **Fe-1** NPs modifies the kinetics of the carbidization process and that the **FeC-1** NPs present different surface properties and tend to agglomerate. Indeed, XPS spectrometry revealed that NPs **FeC-2** were free of Si, contrary to **FeC-1**. XPS also showed that NPs **FeC-2** were richer in Fe(0) at their surface and less oxidized than NPs **FeC-1**, which may be again linked to the presence of the Si containing layer in the latter (see Fig S6).

In order to re-disperse the agglomerates, a sample of 50 mg of the **FeC-1** NPs was heated at 150 °C in mesitylene in the presence of 0.2 equivalents of palmitic acid for 2 hours. A dark-brown suspension that contained the **FeC-3** NPs was recovered after the reaction. Although the suspension was stable for 1 hour and only thereafter the NPs started to precipitate, large agglomerates were still observable by TEM (see Figure S12) together with few isolated NPs. and The **FeC-3** NPs displayed a very low SAR value of ca. 20 W.g$^{-1}$ in hyperthermia experiments at $\mu_0H_{rms}$ = 47 mT and $f$ = 93 kHz. The **FeC-3** NPs were then characterized by VSM and XRD without showing significant differences from **FeC-1** and **FeC-2** NPs (see Fig. S13-15). Moreover, the hysteresis cycles for NPs **FeC-3** and **FeC-2** are identical. Then, the **FeC-3** NPs were further characterised by HAADF-STEM and EDX. No significant differences in the surface composition of the **FeC-3** NPs and the heating **FeC-2** NPs were found by EDX mapping. However, upon observing TEM micrographs at low magnification, it is difficult to determine if the agglomeration degree of the two samples **FeC-2** and **FeC-3** is significantly different (see Figure S14).

*Understanding the heating mechanism of FeC NPs in solution.*

Thanh and co-workers have observed a correlation between the heating properties of multi-core iron oxide NPs and their dispersion in solution.[27] The authors concluded that a better magnetic heating power was associated to less inter-particle interactions as deduced by Dynamic Light Scattering (DLS). Thus, DLS measurements in mesitylene solution evidence for **FeC-3** the presence of large agglomerates displaying a $D_H$ of 590 ± 380 nm (38% population) and 121 ± 40 nm (62% population) (see Figure 2). In contrast, for **FeC-2** displaying a good heating power, a single distribution was observed in the DLS with a $D_H$ of 14.3 ± 1.0 nm. This indicates that inter-particle interactions in solution are a determining factor for explaining the heating properties as previously proposed.

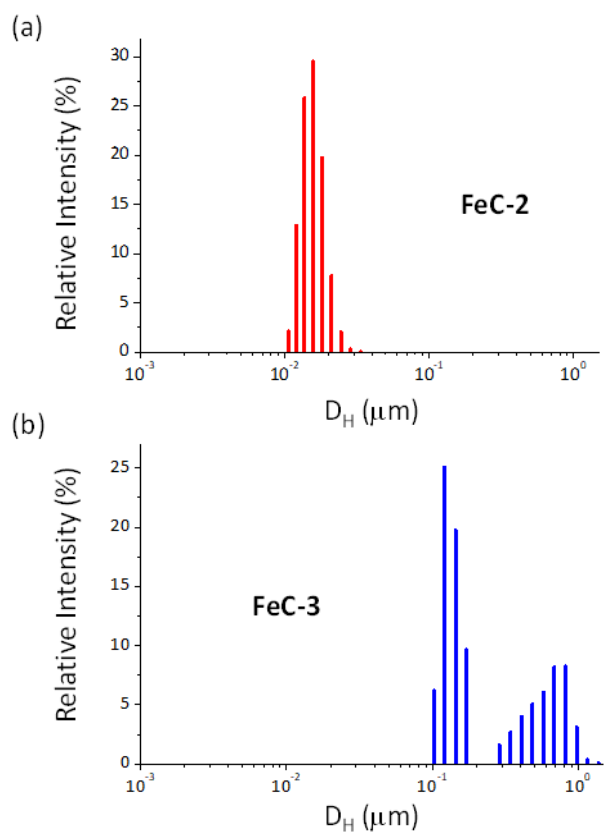

**Figure 2**. Size distribution of the hydrodynamic diameters measured by DLS for the NPs **FeC-2** (a) and **FeC-3** (b).

In order to confirm this hypothesis, we studied the intrinsic magnetic properties of **FeC-2** and **FeC-3** by dispersing the nanoparticles in tetracosane at 60 °C and performing magnetic measurements. Tetracosane solidifies at low temperature and freezes the position of the particles in the medium. This then limits the magnetic interactions between the particles (see Table 2). The **FeC-1** NPs could not be studied as they were not dispersible due to the presence of the larger agglomerates. A zero-field cooled – field cooled (ZFC/FC) experiment with a $\mu_0H_0$ of 10 mT was run, together with the hysteresis loops measurements at 300 K and 5 K (see Fig S16-18). Even though the NPs were diluted in tetracosane to very low Fe concentrations (Fe wt% of ~ 0.01%), in both cases the NPs were far from being well-dispersed and strong interactions between the particles still remained. Thus, the apparent superparamagnetic blocking temperature ($T_b$) at which the ZFC and FC graphs matched was above 300 K, which we attribute to the agglomerates present at different degree in each sample.[48] Higher values of $H_c$ and $M_r$, as well as a saturation of the loops at lower amplitudes of magnetic field were found for **FeC-2** in the hysteresis loops at 5 K after dilution in tetracosane, which agrees with weaker interactions in this sample. ZCF/FC curves indicate that both samples are aggregated but no qualitative conclusion on the difference between the two samples can be provided.

To gain a more precise understanding of the magnetic properties of this system, we turned to measurements under dynamic conditions which are more representative of the situation under hyperthermia conditions. We measured the high-frequency hysteresis loops of both samples under an alternating magnetic field of $\mu_0 H_{rms}$ = 37 mT and 50 kHz using a previously described set-up.[49] The high-frequency hysteresis cycles for the **FeC-2** and **FeC-3** were completely different (see Fig S19): **FeC-2** displayed a fully opened hysteresis loop with a maximum magnetization at 37 mT of 84 A·m$^2$·kg$^{-1}$, whereas **FeC-3** displayed a very small response to the magnetic excitation, with an almost reversible response and a maximum magnetization at 37 mT of 28 A·m$^2$·kg$^{-1}$. The hysteresis area is 23 times larger for **FeC-2** than for **FeC-3**, in agreement with its much more efficient heating. These results demonstrate that although the magnetic properties of the NPs under static conditions are similar for both systems, their dynamic magnetic properties are not comparable.

**Table 3**. Comparison between the heating NPs **FeC-2** and non-heating NPs **FeC-3**.

| Properties | NPs FeC-2 | NPs FeC-3 |
|---|---|---|
| Size (nm) | 14.4 ± 1.3 | 12.1 ± 1.1 |
| SAR (W·g$^{-1}$)[a] | 1260 | 20 |
| $M_s$: (A·m$^2$·kg$^{-1}$) $H_c$: (T) [b] | $M_s$: 170; $H_c$: 0.051 | $M_s$: 168; $H_c$: 0.055 |
| $M_s$: (A·m$^2$·kg$^{-1}$) $H_c$: (T) in tetracosane[c] | $M_s$ ~ 151; $H_c$: 0.020 | $M_s$ ~ 135; $H_c$: 0.025 |
| $M_{max}$: (A·m$^2$·kg$^{-1}$) $H_c$: (T) at $\mu_0H_{rms}$ of 37 mT and $f$ of 50 kHz[d] | $M_{max}$ ~ 84; $H_c$: 0.037 | $M_{max}$ ~ 28; $H_c$: 0.005 |
| $D_H$ by DLS (nm) | 14.3 ± 1.0 | 590 ± 380 (38% in number) 121 ± 40 (62% in number) |

[a] Measured at 47 mT and 93 kHz. [b] $M_s$: Saturation magn. $H_c$: Coercive field. Measured by VSM at 3T and 300 K. [c] $M_s$: Saturation magn. $H_c$: Coercive field. Measured by VSM at 3 T and 300 K after dilution in tetracosane (Fe wt% of ~ 0.01%). [d] $M_{max}$: Magnetization at maximum magnetic field ($\mu_0H_{rms}$ of 37 mT). $H_c$: Coercive field. Cycles were measured in solution at 300 K (10 mg of Fe$_{2.2}$C NPs in 0.5 mL of Mesitylene).

To have a deeper insight into the origin of the difference in heating between the two samples, high-frequency hysteresis loops for **FeC-2** and **FeC-3** were recorded as a function of time using the same set-up. A very different behaviour was found between the two samples (see Fig 3). Whereas the hysteresis loops for NPs **FeC-2** open with time, reaching their maximum value of area after 60 seconds, the hysteresis cycles of the non-heating NPs **FeC-3** remained closed. Fig. S20 displays the evolution of the hysteresis area as a function of time for the two samples. Interestingly, the initial heating power of both samples is quite small and comparable, since the hysteresis area is approximately 0.41 and 0.14 J·kg$^{-1}$ for **FeC-2** and **FeC-3** respectively. However, in the case of **FeC-2**, this value increases up to 7.23 J·kg$^{-1}$ after 60 s without significant evolution afterwards, whereas it remains in the same range for **FeC-3**. Similarly, Fig S21 displays the values of $M_{max}$ and normalized $\chi_m$ as a function of time for NPs **FeC-2** and **FeC-3**. Whereas the $M_{max}$ and $\chi_m$ values remain almost constant for NPs **FeC-3**, $M_{max}$ values get increased by a factor of 7 and $\chi_m$ values by a factor of 10 for NPs **FeC-2** after 60s of exposure to the alternating magnetic field. This is the clear sign that, in sample **FeC-2**, the NPs organize into chains which can be macroscopically observed with a naked eye. This organization induces an effective uniaxial anisotropy along the magnetic field direction, which enhances the heating power of the material. In addition, the area of the high-frequency hysteresis loop after 10 s of exposure has a value of 4.04 J·kg$^{-1}$ (see also Fig S20), which is in of the same order of magnitude as the value measured by calorimetry after the same time of exposure at $\mu_0 H_{rms}$ of 33 mT and a $f$ of 93 kHz (5.67 J·kg$^{-1}$). It should be noted that, for experimental reasons, the concentration in mesitylene of NPs for SAR measurements by calorimetry experiments is ca. 20 mg·mL$^{-1}$, whereas in the dynamic hysteresis measurements it is ca. 10 mg·mL$^{-1}$. Qualitatively, we observed the rapid (1-2 s) formation of chains when the experiments were performed at 5 different concentrations of NPs **FeC-2** (from 2 to 40 mg·mL$^{-1}$). The discrepancy between the measurements for **FeC-2** and **FeC-3** might be due to the fact that larger aggregates are not able to modify their structures to form chains, or that, due to their flux closure magnetic states and the larger inter-aggregate distance, they interact less with each other.

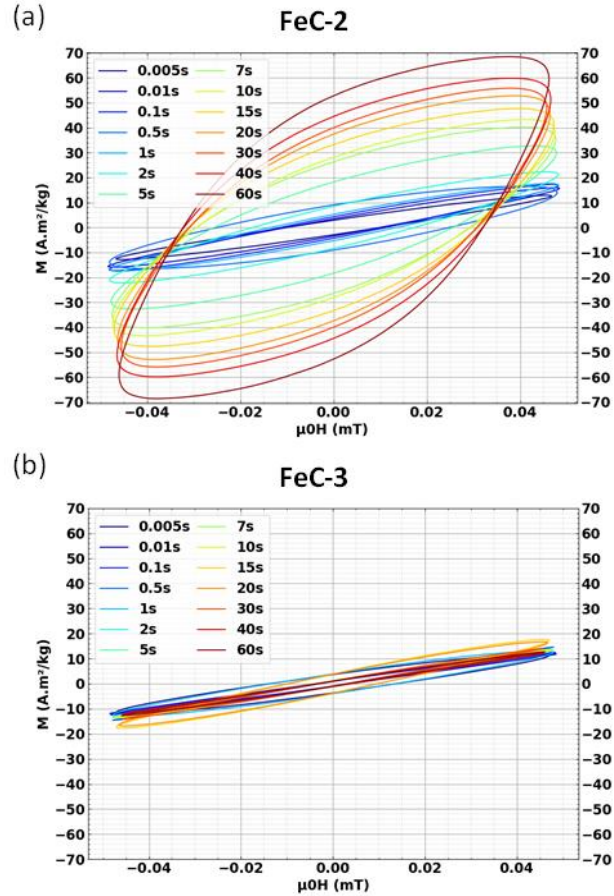

**Figure 3**. High-frequency (50 kHz) hysteresis loops as a function of time for the heating NPs **FeC-2** (a) and for the non-heating NPs **FeC-3** (b). Magnetic field is given by $\mu_0 H_{max}$ ($H_{max} = \sqrt{2} H_{rms}$). A $\mu_0 H_{max}$ of 47 mT in the graph corresponds to a $\mu_0 H_{rms}$ of 33 mT.

To the best of our knowledge, this is the first time that high-frequency hysteresis loops as a function of time are measured for characterizing the heating power of MNPs. First, it clearly evidences that heating power can be a time-dependant property, which was overlooked in previous works. Second, the results obtained on these two samples illustrate that time-dependent high-frequency hysteresis loops constitute a precious tool to understand the origin of heating power in solution of MNPs: whereas the two samples we had measured displayed similar magnetic properties under VSM, their heating properties were very different, which is related to the possibility for one of them to organize in solution.

Finally, we carried out an additional experiment to observe the formation of chains under hyperthermia conditions. When a drop of the above-mentioned solutions was deposed on a grid in the presence of an alternating magnetic field of $\mu_0 H_{rms}$ = 33 mT and 93 kHz and left to evaporate to dryness, the formation of chains of size within the range of 0.5-5 μm that were aligned with the magnetic field direction was clearly observable in the TEM micrographs of **FeC-2** (see Fig 4. and Fig S22). However, in the case of the non-heating NPs **FeC-3**, no anisotropic orientation along the magnetic field could be observed and the NPs remained in the form of big amorphous agglomerates.

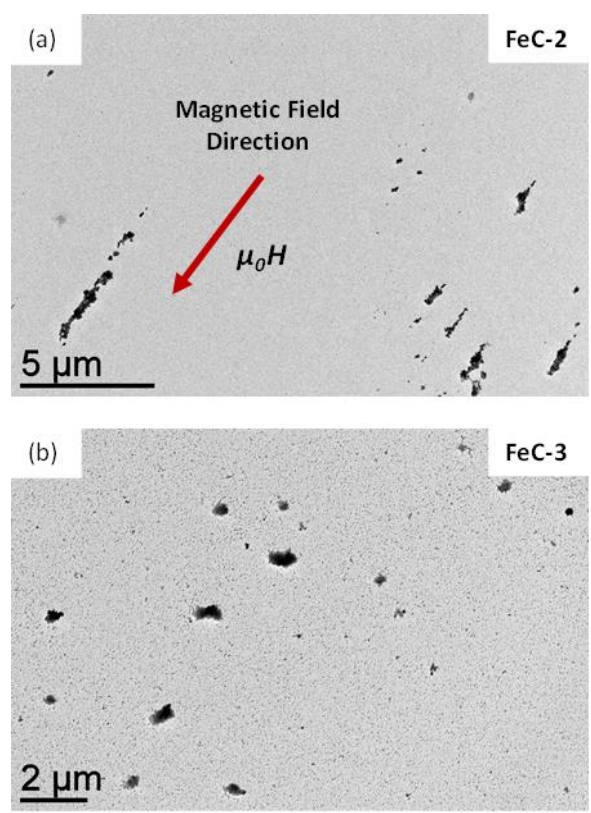

**Figure 4**. TEM micrographs of the NPs (a) **FeC-2** and (b) **FeC-3** when a drop of the hyperthermia solution was deposed in the TEM grid in the presence of an alternating magnetic field. The images illustrate the orientation of the **FeC-2** NPs with the magnetic field, which is responsible of their SAR. Contrary, **FeC-3** NPs remain as amorphous agglomerates.

We conclude this part by presenting a summary of our view on the mechanism at the origin of heating in these samples. The overall hysteresis loop of assemblies of MNPs is the combined result of i) their physical rotation under the influence of the torque or of thermal agitation, and ii) their magnetization reversal. In our case, we consider that agglomerates of MNPs do not rotate under the influence of the magnetic field because they are rather large objects with a very weak remnant magnetization so the torque they undergone is extremely small. In the case of chains, however, we cannot exclude that part of the reversal is due to their physical rotations, but there is so far no suitable model permitting to easily calculate the reversal time of an anisotropic assembly of nanoparticles. Moreover, even if this model existed, it would require knowing precisely the length and width of the chains during the application of the magnetic field. In any case, whatever the underlying mechanism, Fig. 3(a) illustrates well that the coercive field of the chains is close but below the applied alternating magnetic field, which explains why FeC particles display large heating power upon magnetic moment reversal.

*Ligand effects on the heating power of $Fe_{2.2}C$ NPs: reproducible and scalable preparation of nanoparticles displaying high heating power*

The previous results demonstrate that the initial ligand concentration seems to play a crucial role in the carbidization process. This concentration depends on the washing process of the nanoparticles. In this work, we have tried to remove as much ligands as possible before reacting the particles with syngas. This led to **Fe-1** NPs with a Fe concentration of 70-80 wt%. Starting from this sample, and in order to slow down the carbidization process in a controlled way avoiding at the same time the formation of the agglomerates observed in NPs **FeC-1**, a known amount of ligands was added to the NP suspension in mesitylene. In this way, the relation between the nature of the ligands and their concentration in the reaction medium with the SAR of the FeC NPs obtained could be studied.

Four different carbidization experiments were carried out (see experimental section for details), starting from the NPs **Fe-1** with a Fe content of 73 wt% determined by TGA. In these experiments, various amounts of a 1:1 (by weight) mixture of hexadecylamine (HDA) and palmitic acid (PA), the ligands used for the synthesis of the Fe(0) NPs, were introduced into the reactor before reacting with *syngas*. The resulting NPs **FeC-4** to **FeC-7** were thoroughly characterized and their characteristics are summarized in Table 3 (see also Figures S23-28). When carrying out the reaction after addition of solubilizing ligands, all FeC NPs samples could be dispersed in mesytilene and no large agglomerate was observed on the TEM images. In all cases the resulting NPs displayed heating properties in mesitylene solution and a correlation between the SAR and the amount of ligands added before the carbidization process was established. The SAR of the NPs progressively increased when increasing the amount of the 1:1 HDA/PA mixture (see Figure 5). A maximum value of 1700 $W·g^{-1}$ at a field of $\mu_0H_{rms}$ of 47 mT with a frequency of 93 kHz was obtained for NPs **FeC-6** when the reaction was performed starting from a solid sample displaying a 37 wt% Fe content. Further lowering the Fe content down to 24 wt% led to a decrease of the SAR value of NPs **FeC-7** to 670 $W·g^{-1}$. It is also interesting to note that *syngas* consumption at the end of the reaction progressively decreased from **FeC-4** to **FeC-7**, meaning that CO hydrogenation is slowed down when the amount of ligands present in the reaction medium was increased. This observation can be related to the surface accessibility of the nanoparticles. In addition, the crystallite sizes of the $Fe_{2.2}C$ phase determined by the Scherrer equation from the XRD were slightly larger upon increasing the amount of ligands. In all cases, extending the carbidization time to 6 or 7 days did not lead to any notable difference in the heating power of the FeC NPs.

Then, two carbidization experiments were performed under the reaction conditions in which the highest SAR was obtained, i.e. starting from a Fe content of 37%, but adding only PA (**FeC-8**) or HDA (**FeC-9**) as ligands. NPs **FeC-8** and **FeC-9** were characterized by all the above-mentioned techniques (see Fig. S29-33). The nature of the ligand was shown to have a strong influence on the CO hydrogenation rate as deduced from the pressure drop in both experiments at the end of reaction. More precisely, the pressure drop observed was 0.7 bar for **FeC-8** and of >1.5 bar for **FeC-9**, which can be related to the stronger coordination of PA to the NPs as compared to the HDA. In addition, PA was found to produce FeC NPs displaying higher heating power. Thus, the SAR measured for the **FeC-8** NPs was 2020 W·g$^{-1}$ whereas it was only 610 W·g$^{-1}$ for the **FeC-9** NPs when applying a field of $\mu_0H_{rms}$ of 47 mT and a frequency of 93 kHz (see Figure 5 and Experimental section).

The agglomeration degree was again found to play a role on the heating power, since the **FeC-9** NPs were much more agglomerated than the **FeC-8** NPs in agreement with the much better coordination properties of PA compared to HDA. In addition, on the XRD diffractogram of the **FeC-9** NPs, some peaks corresponding to iron oxide were observed. This can be attributed to the competition between water resulting from the carbidization process and the amine ligand for iron coordination, competition which is not present when PA is used. NPs **FeC-8**, were also studied by Mossbauer spectroscopy (see Fig S31). The percentages of the $Fe_{2.2}C$ (69.6%) and $Fe_5C_2$ (24.8%) phases were now comparable to those obtained by Bordet et al. under similar conditions (71.2% of $Fe_{2.2}C$ and 24.4% of $Fe_5C_2$), which displayed also a comparable value of SAR.

Finally, high-frequency hysteresis loops as a function of time were acquired for NPs **FeC-5** and **FeC-8** (see Fig S34-37) at $\mu_0H_{rms}$ of 33 mT and a $f$ of 50 kHz at a concentration of ca. 10 mg·mL$^{-1}$. The area of the cycles for NPs **FeC-5** after 40-60 s were comparable to those found for NPs **FeC-2** (ca. 6-7 mJ·g$^{-1}$), which agrees with their similar values of SAR obtained by calorimetry for these samples. The dynamic of chain formation was also comparable, and the hysteresis loops began to open after 1-2 s of exposure to the alternating magnetic field. However, for the NPs **FeC-8** displaying the highest value of SAR, the area of the hysteresis loops reached a maximum of 14 mJ·g$^{-1}$ after 5-10 s of field exposure, but rapidly decreased because the chains were coming out of the solution as observed in the calorimetry experiments and thus their heating power could not be measured anymore (see Fig S38). Furthermore, chain formation was faster in this case, and the hysteresis loops began to open very rapidly, 0.1 s of exposure to the magnetic field.

**Table 3**. Comparison between the characterizations of NPs **FeC-4** to **FeC-9** prepared by carbidization of the NPs **Fe-1** after addition of different amounts of a 1:1 mixture in mass of HDA/PA, PA or HDA.

| FeC NPs | FeC-4 | FeC-5 | FeC-6 | FeC-7 | FeC-8 | FeC-9 |
|---|---|---|---|---|---|---|
| **Wt% of Fe in starting solid**[a] | 52% | 46% | 37% | 25% | 37% (PA) | 37% (HDA) |
| **ΔP after reaction (bar)** | 1.6 | 1.2 | 0.8 | 0.5 | 0.7 | 2.0 |
| **Size (nm)** | 13.7 ± 1.4 nm | 13.8 ± 1.4 nm | 14.9 ± 1.5 nm | 12.6 ± 1.5 nm | 14.6 ± 1.3 nm | 13.7 ± 2.0 nm |
| **XRD crystallite Size (nm)** | 9.3 | 9.9 | 10.2 | 10.7 | 9.9 | 8.6 (Oxidized) |
| **SAR (W·g$^{-1}$)**[b] | 970 | 1130 | 1700 | 670 | 2020 | 610 |
| **Mössbauer spectroscopy** | -. | Fe$_{2.2}$C (56%) Fe$_5$C$_2$ (29%) Fe(0) (5%) | - | - | Fe$_{2.2}$C (70%) Fe$_5$C$_2$ (25%) Fe(0) (2%) | - |

[a] After addition of a 1:1 PA/HDA mixture in **FeC-4** to **FeC-7**, PA in **FeC-8** or HAD in **FeC-9**. [b] Measured at $\mu_0 H_{rms}$ of 47 mT and a $f$ of 93 kHz.

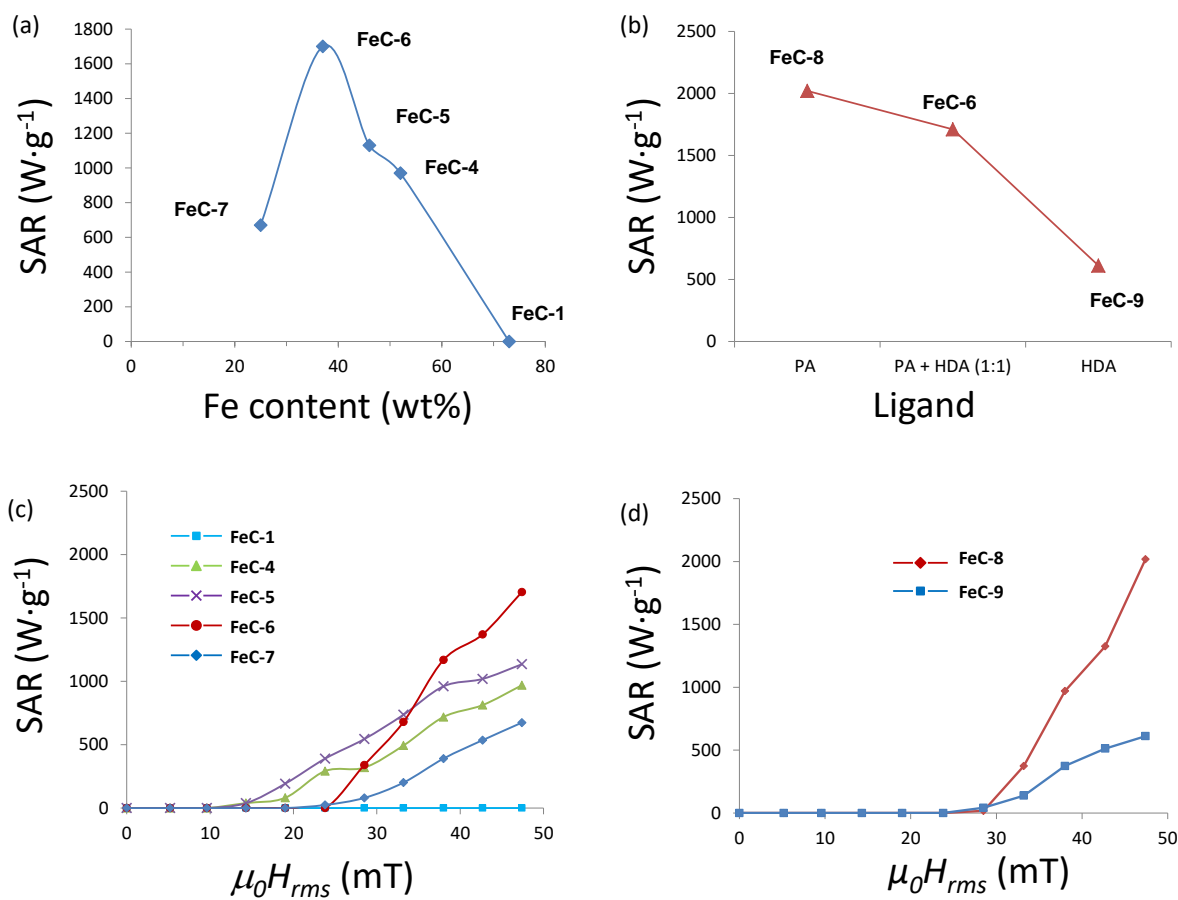

**Figure 5**. (a) Graphical representation of the SAR measured at a field of $\mu_0H_{rms}$ of 47 mT with a frequency of 93 kHz, as a function of the Fe content of the parent Fe(0) NPs. **FeC-1** (no ligands added) NPs **FeC-4** to **FeC-7** (different amounts of a mixture of 1:1 by weight of HDA/PA were added). (b) Graphical representation of the SAR as a function of ligands mixture for NPs **FeC-6**, **FeC-8** and **FeC-9** (when PA:HDA, PA or HDA were respectively used). (c) Comparison of the heating power of the NPs **FeC-1** and **FeC-4** to **FeC-7** at 93 kHz and (d) comparison of the heating power of the NPs **FeC-8** and **FeC-9** at 93 kHz

All these experiments show that, in order to obtain FeC NPs displaying a good heating power, the presence of an optimum quantity of ligands is necessary (ca. 100 wt% respect to the Fe(0) NPs). The role that the ligands play in the reaction mechanism is to prevent the agglomeration of the NPs in solution and to slow down the carbon incorporation rate (see Scheme 1).

The mobility in solution of well-dispersed FeC NPs is an important factor to obtain high SAR values. The nature of the ligand also influences the preparation of FeC NPs. PA coordinates more strongly to the surface of the Fe(0) NPs than HDA.[42] The oxidation of the NPs **FeC-9** observed in the XRD is in agreement with this hypothesis. If HDA gives weaker interactions with the NPs, they are more prone to react with the water that is generated in the carbidization process leading to partial oxidation of the surface. In addition, PA may assist the shift of the surface-coordinated *hmds* by protonation, hence helping to avoid the formation of the Si containing amorphous layer that embeds the agglomerates composing the non-heating NPs **FeC-1**. Understanding the influence of the reaction conditions allowed preparing reproductively and in a scalable way NPs **FeC-8**, which display the highest heating power in this work (ca. 2000 W·g$^{-1}$) at a field of $\mu_0H_{rms}$ of 47 mT with a *f* of 93 kHz.

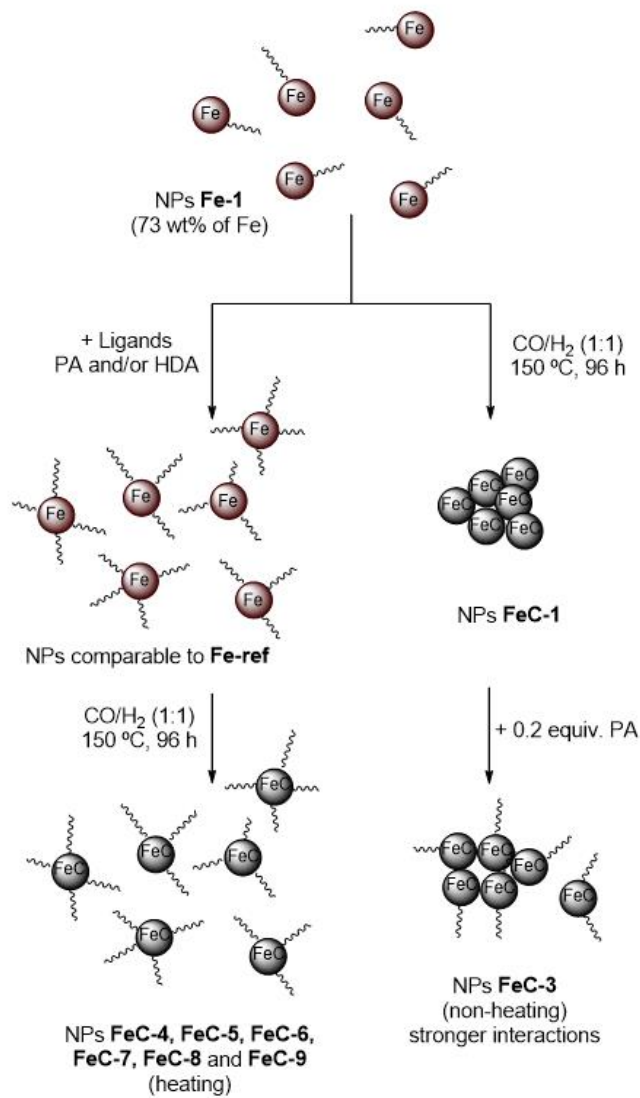

**Scheme 1**. Schematic representation of the role of the ligands into the heating properties of the Fe$_{2.2}$C NPs.

CONCLUSIONS

In this work, in order to scale-up the reproducible synthesis of highly heating magnetic nanoparticles and to understand their heating mechanism, we have extended our previous studies concerning the synthesis under *syngas* of FeC NPs by carbidization of pre-formed Fe(0) NPs. We have studied their structure and their heating properties in detail. We have discovered a surprising effect of the extended washing of the starting Fe(0) nanoparticles. When the nanoparticles surface is depleted of its ligands after extended washing, the particles are strongly and irreversibly agglomerated and cannot align in an alternating magnetic field. The result is the absence of heating properties of these nanomaterials. Control over the ligands concentration on the surface of the NPs is therefore a pre-requisite to prepare nanoparticles displaying a controllable and reproducible heating power. In parallel, this work explores the heating mechanism in magnetic nanoparticles, demonstrates the importance of dynamic magnetic measurements and brings forward a new indication of the importance of dipolar interactions, here specifically in solution, for controlling the heating power of magnetic materials.

Finally, when controlling the Fe concentration at the beginning of the reaction, the carbidization process showed good reproducibility and scalability. We believe that this work represents an important advance in the synthetic control of well-dispersed homogeneous magnetic NPs, here FeC NPs, displaying reliable heating properties in the objective of future applications.

CONCFLICTS OF INTEREST

The authors declare no conflict of interest

AKNOWLEDGMENTS

The authors thank ERC Advanced Grant (MONACAT 2015-694159) for financial support, J.-F. Meunier for Mössbauer measurements and Jérôme Esvan and Simon Tricard for the XPS measurements.

NOTES AND REFERENCES

1. D. H. Ortgies, F. J. Teran, U. Rocha, L. de la Cueva, G. Salas, D. Cabrera, A. S. Vanetsev, M. Rähn, V. Sammelselg, Y. V. Orlovskii and D. Jaque, *Advanced Functional Materials*, 1704434-n/a.

2.	J. Yu, F. Chen, W. Gao, Y. Ju, X. Chu, S. Che, F. Sheng and Y. Hou, *Nanoscale Horiz.*, 2017, **2**, 81-88.

3.	E. A. Perigo, G. Hemery, O. Sandre, D. Ortega, E. Garaio, F. Plazaola and F. J. Teran, *Appl. Phys. Rev.*, 2015, **2**, 041302/041301-041302/041335.

4.	A. Meffre, B. Mehdaoui, V. Connord, J. Carrey, P. F. Fazzini, S. Lachaize, M. Respaud and B. Chaudret, *Nano Lett.*, 2015, **15**, 3241-3248.

5.	A. Bordet, L.-M. Lacroix, P.-F. Fazzini, J. Carrey, K. Soulantica and B. Chaudret, *Angew. Chem., Int. Ed.*, 2016, **55**, 15894-15898.

6.	S. Ceylan, C. Friese, C. Lammel, K. Mazac and A. Kirschning, *Angew. Chem., Int. Ed.*, 2008, **47**, 8950-8953.

7.	S. Ceylan, L. Coutable, J. Wegner and A. Kirschning, *Chem. - Eur. J.*, 2011, **17**, 1884-1893, S1884/1881-S1884/1818.

8.	J. Hartwig, S. Ceylan, L. Kupracz, L. Coutable and A. Kirschning, *Angew. Chem., Int. Ed.*, 2013, **52**, 9813-9817.

9.	P. M. Mortensen, J. S. Engbaek, S. B. Vendelbo, M. F. Hansen and M. Oestberg, *Ind. Eng. Chem. Res.*, 2017, **56**, 14006-14013.

10.	WO2014162099A1, 2014.

11.	M. G. Vinum, M. R. Almind, J. S. Engbaek, S. B. Vendelbo, M. F. Hansen, C. Frandsen, J. Bendix and P. M. Mortensen, *Angew. Chem., Int. Ed.*, 2018, **57**, 10569-10573.

12.	A. B. Salunkhe, V. M. Khot and S. H. Pawar, *Curr. Top. Med. Chem.*, 2014, **14**, 572-594.

13. N. A. Usov, *J. Appl. Phys.*, 2010, **107**, 123909/123901-123909/123912.

14. J. Carrey, B. Mehdaoui and M. Respaud, *J. Appl. Phys.*, 2011, **109**, 083921/083921-083921/083917.

15. C. L. Dennis, A. J. Jackson, J. A. Borchers, R. Ivkov, A. R. Foreman, J. W. Lau, E. Goernitz and C. Gruettner, *J. Appl. Phys.*, 2008, **103**, 07A319/311-307A319/313.

16. F. Burrows, C. Parker, R. F. L. Evans, Y. Hancock, O. Hovorka and R. W. Chantrell, *J. Phys. D: Appl. Phys.*, 2010, **43**, 474010/474011-474010/474010.

17. C. L. Dennis, A. J. Jackson, J. A. Borchers, P. J. Hoopes, R. Strawbridge, A. R. Foreman, J. van Lierop, C. Gruttner and R. Ivkov, *Nanotechnology*, 2009, **20**, 395103/395101-395103/395107.

18. C. Haase and U. Nowak, *Phys. Rev. B: Condens. Matter Mater. Phys.*, 2012, **85**, 045435/045431-045435/045435.

19. B. Mehdaoui, R. P. Tan, A. Meffre, J. Carrey, S. Lachaize, B. Chaudret and M. Respaud, *Phys. Rev. B: Condens. Mater Mater. Phys.*, 2013, **87**, 174419/174411-174419/174410.

20. C. Martinez-Boubeta, K. Simeonidis, D. Serantes, I. Conde-Leboran, I. Kazakis, G. Stefanou, L. Pena, R. Galceran, L. Balcells, C. Monty, D. Baldomir, M. Mitrakas and M. Angelakeris, *Adv. Funct. Mater.*, 2012, **22**, 3737-3744, S3737/3731-S3737/3738.

21. R. P. Tan, J. Carrey and M. Respaud, *Phys. Rev. B: Condens. Matter Mater. Phys.*, 2014, **90**, 214421/214421-214421/214412, 214412 pp.

22. D. Serantes, K. Simeonidis, M. Angelakeris, O. Chubykalo-Fesenko, M. Marciello, M. d. P. Morales, D. Baldomir and C. Martinez-Boubeta, *The Journal of Physical Chemistry C*, 2014, **118**, 5927-5934.

23. E. Myrovali, N. Maniotis, A. Makridis, A. Terzopoulou, V. Ntomprougkidis, K. Simeonidis, D. Sakellari, O. Kalogirou, T. Samaras, R. Salikhov, M. Spasova, M. Farle, U. Wiedwald and M. Angelakeris, *Sci. Rep.*, 2016, **6**, 37934.

24. K. Simeonidis, M. P. Morales, M. Marciello, M. Angelakeris, P. de la Presa, A. Lazaro-Carrillo, A. Tabero, A. Villanueva, O. Chubykalo-Fesenko and D. Serantes, *Sci. Rep.*, 2016, **6**, 38382.

25. J. G. Ovejero, D. Cabrera, J. Carrey, T. Valdivielso, G. Salas and F. J. Teran, *Phys. Chem. Chem. Phys.*, 2016, **18**, 10954-10963.

26. R. Das, J. Alonso, Z. Nemati Porshokouh, V. Kalappattil, D. Torres, M.-H. Phan, E. Garaio, J. A. Garcia, J. L. Sanchez Llamazares and H. Srikanth, *J. Phys. Chem. C*, 2016, **120**, 10086-10093.

27. C. Blanco-Andujar, D. Ortega, P. Southern, Q. A. Pankhurst and N. T. K. Thanh, *Nanoscale*, 2015, **7**, 1768-1775.

28. F. M. Martin-Saavedra, E. Ruiz-Hernandez, A. Bore, D. Arcos, M. Vallet-Regi and N. Vilaboa, *Acta Biomater.*, 2010, **6**, 4522-4531.

29. Z. Li, M. Kawashita, N. Araki, M. Mitsumori, M. Hiraoka and M. Doi, *J. Biomater. Appl.*, 2011, **25**, 643-661.

30. E. M. Muzquiz-Ramos, V. Guerrero-Chavez, B. I. Macias-Martinez, C. M. Lopez-Badillo and L. A. Garcia-Cerda, *Ceram. Int.*, 2015, **41**, 397-402.

31. P. Drake, H.-J. Cho, P.-S. Shih, C.-H. Kao, K.-F. Lee, C.-H. Kuo, X.-Z. Lin and Y.-J. Lin, *J. Mater. Chem.*, 2007, **17**, 4914-4918.

32. K. H. Bae, M. Park, M. J. Do, N. Lee, J. H. Ryu, G. W. Kim, C. Kim, T. G. Park and T. Hyeon, *ACS Nano*, 2012, **6**, 5266-5273.

33. P. Guardia, R. Di Corato, L. Lartigue, C. Wilhelm, A. Espinosa, M. Garcia-Hernandez, F. Gazeau, L. Manna and T. Pellegrino, *ACS Nano*, 2012, **6**, 3080-3091.

34. A. Meffre, B. Mehdaoui, V. Kelsen, P. F. Fazzini, J. Carrey, S. Lachaize, M. Respaud and B. Chaudret, *Nano Lett.*, 2012, **12**, 4722-4728.

35. C. Yang, H. Zhao, Y. Hou and D. Ma, *J. Am. Chem. Soc.*, 2012, **134**, 15814-15821.

36. Z. Yang, T. Zhao, X. Huang, X. Chu, T. Tang, Y. Ju, Q. Wang, Y. Hou and S. Gao, *Chem. Sci.*, 2017, **8**, 473-481.

37. Z.-Y. Wu, X.-X. Xu, B.-C. Hu, H.-W. Liang, Y. Lin, L.-F. Chen and S.-H. Yu, *Angew. Chem. Int. Ed.*, 2015, **54**, 8179-8183.

38. S. Y. Hong, D. H. Chun, J.-I. Yang, H. Jung, H.-T. Lee, S. Hong, S. Jang, J. T. Lim, C. S. Kim and J. C. Park, *Nanoscale*, 2015, **7**, 16616-16620.

39. K. Xu, B. Sun, J. Lin, W. Wen, Y. Pei, S. Yan, M. Qiao, X. Zhang and B. Zong, *Nat. Commun.*, 2014, **5**, 5783.

40. F. Dumestre, B. Chaudret, C. Amiens, P. Renaud and P. Fejes, *Science*, 2004, **303**, 821.

41. C. Desvaux, C. Amiens, P. Fejes, P. Renaud, M. Respaud, P. Lecante, E. Snoeck and B. Chaudret, *Nat. Mater.*, 2005, **4**, 750-753.

42. L.-M. Lacroix, S. Lachaize, A. Falqui, M. Respaud and B. Chaudret, *J. Am. Chem. Soc.*, 2009, **131**, 549-557.

43. A. Meffre, S. Lachaize, C. Gatel, M. Respaud and B. Chaudret, *J. Mater. Chem.*, 2011, **21**, 13464-13469.

44. A. Bordet, L.-M. Lacroix, K. Soulantica and B. Chaudret, *ChemCatChem*, 2016, **8**, 1727-1731.

45. B. Mehdaoui, A. Meffre, L. M. Lacroix, J. Carrey, S. Lachaize, M. Gougeon, M. Respaud and B. Chaudret, *J. Magn. Magn. Mater.*, 2010, **322**, L49-L52.

46. Y. Sun, Q. Wang, Z. Qu and M. Shi, *Key Eng. Mater.*, 2012, **512-515**, 1424-1428.

47. X. Batlle, M. G. Del Muro and A. Labarta, *Phys. Rev. B: Condens. Matter*, 1997, **55**, 6440-6445.

48. M. Mikhaylova, D. K. Kim, N. Bobrysheva, M. Osmolowsky, V. Semenov, T. Tsakalakos and M. Muhammed, *Langmuir*, 2004, **20**, 2472-2477.

49. V. Connord, B. Mehdaoui, R. P. Tan, J. Carrey and M. Respaud, *Rev. Sci. Instrum.*, 2014, **85**, 093904/093901-093904/093908.